\documentclass[aip,jcp,reprint,showkeys,superscriptaddress,longbibliography]{revtex4-2}
\usepackage[utf8]{inputenc}

\usepackage[T1]{fontenc}
\usepackage{amsmath,amssymb,mathtools,mleftright}
\usepackage{txfonts}

\usepackage{mathrsfs}

\usepackage{graphicx,grffile}
\graphicspath{{figures/}}
\newlength{\figwidth}
\usepackage{siunitx}

\usepackage{url,hyperref}
\usepackage[usenames,dvipsnames]{xcolor}
\hypersetup{colorlinks=true, linkcolor=BrickRed, urlcolor=blue!50!black, citecolor=blue!50!black}

\usepackage[capitalize]{cleveref}
\crefrangelabelformat{equation}{(#3#1#4)$-$(#5#2#6)}

\renewcommand\leq{\leqslant}
\renewcommand\geq{\geqslant}
\renewcommand\phi\varphi
\renewcommand\rho\varrho
\renewcommand\vec[1]{\textrm{\bfseries #1}}

\newcommand\diff{\mathrm{d}}

\begin{document}
\title{Critical surface adsorption of confined binary liquids with locally conserved mass and composition}

\author{Sutapa Roy}
\affiliation{Department of Physics, University of Milan, via Celoria 16, Milano 20133, Italy}

\author{Felix Höf{}ling}
\affiliation{Freie Universität Berlin, Department of Mathematics and Computer Science, Arnimallee 6, 14195 Berlin, Germany}
\affiliation{Zuse Institute Berlin, Takustr.\ 7, 14195 Berlin, Germany}

\date{\today}

\begin{abstract}
Close to a solid surface, the properties of a fluid deviate significantly from their bulk values.
In this context, we study the surface adsorption profiles of a symmetric binary liquid confined to a slit pore by means of molecular dynamics simulations; the latter naturally entails that mass and concentration are locally conserved.
Near a bulk consolute point, where the liquid exhibits a demixing transition with the local concentration as the order parameter, we determine the order parameter profiles and characterise the relevant critical scaling behaviour, in the regime of strong surface attraction, for a range of pore widths and temperatures.
The obtained order parameter profiles decay monotonically near the surfaces, also in the presence of a pronounced layering in the number density.
Overall, our results agree qualitatively with recent theoretical predictions from a mesoscopic field-theoretical approach for the canonical ensemble.
\end{abstract}


\maketitle

\section{Introduction}




The investigation of fluids under nano-confinement has drawn substantial research attention in the past decades.
In this context, water doubtlessly belongs to the most prominent liquid substance whereas it is arguably also one of the most difficult one to understand.
To this end, molecular simulations have provided key insights into the properties of water, complementing experimental findings and theoretical analyses (see, e.g., Ref.~\citenum{gallo2021epje} and references therein). This includes the calculation of phase diagrams, e.g., of liquid--vapour and fluid--solid equilibria \cite{vega2006jcp,vega2008jpcm}
and a putative liquid--liquid transition deep in the super-cooled state \cite{abascal2010jcp}, 
but also structural properties \cite{gallo2021epje} and transport coefficients such as the dynamic shear viscosity \cite{dehaoui2015pnas,straube2020cp,schulz2020prf}.

Confining a liquid in channels of only a few nanometers in diameter manifestly breaks the translational symmetry
and can significantly alter its physical properties relative to its behavior in bulk. For example, certain crystal structures can be suppressed and phase transition temperatures can be shifted \cite{berwanger2009}.
Confinement intricately modifies the relaxation dynamics of glass-forming \cite{lang2010prl, paluch2019glass, Arutkin2020glass} and polymeric \cite{Varnik:JCP2002, Richter2013, Richter2019, Riggleman2019} liquids, as well as water again \cite{brovchenko2003epje, falk2010nl}.
The presence of solid surfaces modifies the transversal density profiles showing, e.g., layering,
which can lead to surface-induced phase transitions such as
pre-freezing and pre-melting \cite{eshraghi2018sm} and the rich scenario of wetting and drying transitions \cite{nakanishi1982prl, dietrich1988, finn1989pra, fan1993jcp, binder2003jsp, maciolek2003jcp, brovchenko2005jpcm, brovchenko2007pre, brovchenko2008pre, brovchenko2012, evans2016prl, evans2017jcp, evans2019pnas, parry2023prl}; predictions that long-ranged dispersion forces lead to specific phase diagrams \cite{evans2019pnas} were criticised recently \cite{parry2023prl}.
For complex liquids, consisting of anisotropic molecules, anchoring effects to the surface add another fascinating twist to the wetting problem \cite{rodriguez-ponce1999prl, mueller2003ms, wu2018jcp},
whereas orientational ordering can be observed for rod-like, stiff polymers \cite{nikoubashman2017polymer}.

Going beyond single-component fluids, liquid mixtures exhibit phase separation, which in conjunction with confinement can lead to surface enrichment and domain growth of one component \cite{puri1993jcp,frisch1994,das2006sdsd,das2006pre}.
The demixing transition of the bulk phase diagram is associated with a continuous phase transition and, thus, with the emergence of a macroscopic correlation length and relaxation time \cite{hohenberg1977, folk2006, burstyn1983, sengersbook2014, das2003, das2006jcp, roy2011epl, roy2016, pathania2021ats}.
Furthermore, the transversal pore dimensions delimit the growth of critical fluctuations so that the critical singularities near the consolute point appear rounded \cite{binder2008, basu2016epl}.
It also gives rise to critical Casimir forces \cite{krech1999jpcm, vasilyev2015, fukuto2005prl, rafai2007physa, hertlein2008, dantchev2023pr, Gambassi:SM2024, Wang:NC2024}
and to a splitting of the bulk universality class into few surface universality classes \cite{diehl1997,dietrich1988} (see below).
In particular, for the local excess concentration, playing the role of the order parameter (OP), one expects that its spatial dependence near a solid substrate exhibits critical scaling.
This surface-critical behaviour was corroborated within Monte Carlo simulations for
Ising-like lattice fluids \cite{binder-landau1990, binder2003jsp} and for
single-component Lennard-Jones (LJ) liquids \cite{brovchenko2005epjb, brovchenko2012}, with the local density as the OP field.

Monte Carlo simulations on phase coexistence are typically carried out either in the grand canonical or in the Gibbs ensemble. Also the majority of theoretical studies on critical phenomena in confinement have been performed in the grand canonical ensemble.
Experimentally, however, the mass of each component of a mixture is locally conserved, the corresponding total masses are fixed, and thus the canonical ensemble appears as the proper description of confined liquids in thermal equilibrium.
Whereas the statistical ensembles are equivalent with respect to the mean values of thermodynamic observables, they are not with respect to their fluctuations \cite{plischke, lebowitz1967}.
Further on the level of thermodynamic macrostates, the ensemble equivalence may not hold in the presence of long-ranged interactions (see, e.g., refs.~\citenum{touchette2004physa} and \citenum{touchette2015jsp} and references therein).
This issue becomes particularly relevant for confined fluids close to their critical points and, as a result, the choice of the ensemble can yield very different predictions for the behaviour of the fluid.
For example, the critical Casimir force emerging for a fluid confined in between two parallel walls with symmetric boundary conditions is \emph{repulsive} for the canonical ensemble whereas it is \emph{attractive} for the grand canonical ensemble \cite{gross2016, gross2017, gross2018, rohwer2019pre}.


In the present study, we employ state-of-the-art molecular dynamics (MD) simulations to investigate the surface adsorption of a
confined binary liquid film \textit{at} its bulk demixing (liquid--liquid) critical point.
The binary mixture considered is symmetric with respect to its components \cite{das2003, das2006jcp, roy2016}, and both walls of the slit pore preferentially attract, with equal strength, the same component of the mixture.
Within the MD simulations, as well as in experiments, the partial masses of the two components are naturally conserved
so that the data obtained for equilibrium properties correspond to the canonical ensemble.
Specifically, we calculate spatial profiles of the excess adsorption and analyse their critical scaling.
Our findings are compared with existing mean-field predictions in the limit of infinite surface attraction strengths.
We note that computational studies of near-critical fluids are notoriously challenging due to critical slowing down \cite{folk2006}, a macroscopically large correlation length (entailing finite-size corrections \cite{binder1981, roy2013, roy2014, pathania2021ats}), and strong OP fluctuations (expressed by a high compressibility or generalised susceptibility); we have addressed these issues by performing a large number of independent simulations encompassing long time spans and huge particle numbers.

\section{Critical surface adsorption}

Upon approaching the critical point of a continuous phase transition, the local OP near a boundary deviates in normal direction from its bulk value on the scale of the diverging bulk correlation length \cite{binder_review1983, diehl_review1983}.
The kind of thermodynamic singularities occurring in this surface layer depends on the boundary conditions for the OP such that each bulk universality class splits up into various surface universality classes.
Generically, fluids belong to the so-called normal surface universality class \cite{dietrich1988}, which is characterized by the presence of a symmetry breaking surface field inducing order at the surface even if the bulk is in the disordered phase.
Close to the critical point the bulk correlation length $\xi_b$ diverges in the thermodynamic limit, thus becoming the most predominant length scale in the system.
In the following, we specialise to binary fluids near their bulk consolute point, given by the critical temperature $T_c$ and the
critical composition (which plays a similar role as the critical density in case of a liquid--vapour transition).
In addition, we restrict the discussion to the mixed phase, i.e., to fluid temperatures $T > T_c$,
and we will often incorporate the temperature dependence via the bulk correlation length $\xi_b = \xi_b(T)$.
For this choice of temperatures, wetting transitions are not expected to occur \cite{nakanishi1982prl, binder2003jsp, parry2023prl}.

The demixing OP of a binary liquid of species A and B in bulk is given by $\phi_b=x_A-x_B$ in terms of the bulk concentrations $x_A$ and $x_B$ of both species, defined as their corresponding mole fractions.
Upon confinement to a slit pore, these quantities depend on the perpendicular distance $z$ from one of the confining walls,
so that the local OP reads $\Phi(z)=x_A(z)-x_B(z)$, where $x_\alpha(z) = \rho_\alpha(z) / \rho(z)$ for $\alpha=A,B$;
the partial number density $\rho_\alpha(z)$ counts the particles of species $\alpha$ in a narrow slab at position $z$, and
$\rho(z) = \rho_A(z) + \rho_B(z)$ is the overall local number density.

For fluids in contact with a single wall, $\xi_b$ competes with the range of the surface interaction of the fluid molecules;
in the case of a short ranged interaction, as studied here, this range is typically on the order of the molecular diameter $\sigma$.
A universal shape of the OP profile $\Phi(z)$, as function of the distance $z$ to the surface, is anticipated except very close to the surface, i.e., universality holds for $z \gg \sigma$.
Depending on the distance $z$ compared to the correlation length $\xi_b$, one distinguishes a standard regime with an exponential decay, $\Phi(z) \sim \exp(-z/\xi_b)$ for $z \gg \xi_b$, and a critical regime exhibiting power-law behaviour, $\Phi(z) \sim z^{-\beta/\nu}$ for $z \ll \xi_b$.
The exponents $\beta$ and $\nu$ denote the standard critical exponents of the three-dimensional Ising universality class; we use the reliable estimates \cite{pelissetto2002} $\nu = 0.630$ and $\beta = 0.325$.
In particular, they govern the critical singularities in the bulk behaviour of the correlation length and the OP, respectively \cite{fisher1971,sengersbook2014}:
$\xi_b(T) \simeq \xi_0^+ \epsilon^{-\nu}$ in terms of $\epsilon = (T-T_c)/T_c \downarrow 0$, measuring the proximity to the bulk critical point, and $\phi_b(T) \simeq \phi_0 (-\epsilon)^\beta$ for $\epsilon \uparrow 0$,
where $\xi_0^+$ and $\phi_0$ are non-universal amplitudes.

The OP profile exhibits critical scaling which, for a semi-infinite fluid, is condensed in a dimensionless scaling function $P_1$ with the rescaled distance $\hat z := z/\xi_b$ as the only variable \cite{ciach1990,dietrich1995}:
\begin{equation}\label{opsemiinfinite}
  \Phi(z;T>T_c) = \varphi_0 (\xi_b/\xi_0^+)^{-\beta/\nu} P_1(z/\xi_b) \,, \quad z \gg \sigma \,.
\end{equation}
The scaling function is universal, but depends on the boundary condition for the OP at the surface.
The scaling function obeys the asymptotics sketched above, namely
$P_1(\hat z \to \infty) \simeq \exp(-\hat z)$ and
$P_1(\hat z \to 0) \simeq c_0 \hat z^{-\beta/\nu}$ with the universal surface amplitude $c_0$.
The excess mass of fluid adsorbed to the surface is quantified by the integrated OP profile,
\begin{equation}
 \Gamma(T) = \int_0^\infty [\Phi(z;T) - \phi_b(T)] \, \diff z \,,
\end{equation}
for which, upon inserting \cref{opsemiinfinite}, one infers a critical power-law divergence \cite{dietrich1995}:
\begin{equation} \label{critical-adsorption}
 \Gamma(T \to T_c) \sim \xi_b^{1-\beta/\nu} \sim |\epsilon|^{\beta - \nu} \,,
\end{equation}
provided that $\beta < \nu$.
(In mean-field theories, where $\beta = \nu = 1/2$, a logarithmic divergence is predicted).
These scaling predictions for the OP profile have been tested successfully in simulations of Ising lattices \cite{binder-landau1990}.
Scaling has also been corroborated in simulations of one-component LJ fluids,
but the verification of the critical divergences has remained challenging \cite{brovchenko2008pre, brovchenko2012}.

For fluids confined to a slit pore, surface effects due to both walls are present and, as a result, the distance $D$ between the two walls enters the problem as another relevant length.
In particular, critical fluctuations are suppressed in the direction of the surface normal for $\xi_b \gg D$, whereas the slit confinement is expected to be ineffective for $\xi_b \ll D$, approximating the situation of a semi-infinite fluid delimited by a single surface.
The OP profile $\Phi(z)$ is now a function of the position $z$, the temperature $T$, and the pore width $D$;
it also depends on the strength of the surface interaction, and we may again use the bulk correlation length $\xi_b(T)$ instead of $T$.
The dependence on the pore width $D$ is accounted for by decorating the scaling function in \cref{opsemiinfinite} with a second scaling variable, $\hat D = D/\xi_b$, which is further used to substitute $\xi_b$ by $D$, i.e., we compare other quantities such as the position $z$ to the pore width~$D$.
In the strong adsorption limit (large surface field), the finite-size scaling form of the OP profile reads \cite{diehl1997}:
\begin{equation}\label{op1}
  \Phi(z;T,D) = \varphi_0 \left(\frac{D}{\xi_0^+} \right)^{-\beta/\nu} P_2\mleft({\frac{z}{D}}, \frac{D}{\xi_b} \mright) \,,
  \quad z \gg \sigma \,.
\end{equation}
Thus, plotting
$\varphi_0^{-1}(D/\xi_0^+)^{\beta/\nu} \Phi(z;T,D)$ vs.\ $\bar z = z/D$ for a fixed value of $\hat D = D/\xi_b$ should yield a data collapse.
The new, two-parameter scaling function $P_2$ reproduces the behaviour near a single surface via
$P_1(\hat z) = \lim_{\hat D \to \infty} \hat D^{-\beta/\nu} P_2(\hat z/\hat D, \hat D)$
and it encodes the critical profile $\Phi(z; T_c, D)$ as the limit $P_2(\bar z, \hat D \to 0)$.
The function $P_2$ also depends on the boundary conditions at the two surfaces; here, we focus on a slit pore with equal surfaces on both sides, namely symmetric, attractive boundary conditions.

We note that, due to the confinement, the critical divergence at the bulk critical values of temperature and composition is shifted towards a lower temperature and a more A-rich composition (assuming A-preferring, symmetric surfaces) \cite{nakanishi1982prl, binder2003jsp, das2006sdsd, das2006pre}.
The phenomenon is related to capillary condensation and depends on the pore width $D$; it is particularly relevant for thin films, $D \ll \xi_b$.
More precisely, it may be expressed as a competition of the pore width $D$ and the bulk correlation length $\xi_b$ and can be incorporated in the scaling form \eqref{op1} by adding an additional scaling variable for the composition.
However, we will not explore these aspects here and restrict to $T > T_c$ in the following, where $T_c$ always refers to the critical temperature of the bulk fluid, which also enters \cref{op1} through the scaling of $\xi_b$.

For a conserved OP field, the critical scaling of the OP profiles has been studied in depth recently within a mesoscopic theory based on the Ginzburg--Landau (GL) free energy functional, prescribing either a surface field \cite{gross2016, gross2017} or the contact values at the two surfaces \cite{rohwer2019pre}. 
In these works, analytic calculations within a systematic perturbative approach yield OP scaling functions, which are corroborated by numerical solutions of the field equations;
in addition, these findings are supported qualitatively with data from Monte Carlo simulations for Ising lattices.
The obtained OP profiles are mean-field like, as is typical for GL theories, and suggest
that, for symmetric boundary conditions [see eq.~(33) of Ref.~\citenum{gross2016}],
\begin{equation}
  \Phi(z;T, D) \simeq \mu \xi_b^2 - \Phi_m(T, D) \,
    \cosh\mleft(\frac{z - D/2}{\xi_b} \mright)
  \label{op-cosh-profile}
\end{equation}
in the inner part of the pore, \emph{viz.}, for $\xi_b \ll z \ll D - \xi_b$.
Here, we consider binary fluids at critical composition so that the chemical potential $\mu$ associated with the OP field is zero and the first term vanishes.
\footnote{A similar expression was obtained for Dirichlet boundary conditions at both surfaces [see eq.~(24) of Ref.~\citenum{rohwer2019pre}]. However, the situations studied analytically in Refs.~\citenum{gross2016} and \citenum{rohwer2019pre} differ from the present setup, which corresponds to the boundary conditions $\Phi(0) = \Phi(D) = \Phi_0 \approx 1$ and zero total OP, $\phi_\text{tot} = 0$.}
%
%
%
%
With this, the prefactor $\Phi_m = \Phi(z = D/2)$ equals the minimum value of the OP profile, which is attained in the central plane of the slit pore by symmetry.
Furthermore, the local conservation of the OP field implies a constraint on the total OP $\phi_\text{tot}$, which is the spatial integral of $\Phi(z)$ and which vanishes in the mixed phase at critical composition:
\begin{equation} \label{op-tot}
  \phi_\text{tot} := \int_0^D \Phi(z) \, \diff z = 0 \,.
\end{equation}
Therefore and exploiting the symmetry $\Phi(z) = \Phi(D-z)$ of the set-up, $\Phi_m < 0$ is related to the excess adsorption $\Gamma$ on each surface via
\begin{equation}
  \Gamma(T,D) := \int_0^{D/2} [\Phi(z) - \Phi_m] \, \diff z = -\frac{\Phi_m D}{2} \,.
  \label{phi-m}
\end{equation}

For a finite pore width, $D < \infty$, the profile $\Phi(z; T, D)$ is expected to depend on the temperature $T$ analytically near $T_c$, which implies that the scaling function $P_2(\bar z, \hat D)$ is analytic in $\hat D^{1/\nu}$ with $\hat D = D / \xi_b$ [\cref{op1}].
Hence, $\Phi_m(T, D)$ may be expanded for $T \downarrow T_c$ or, equivalently, for $\xi_b(T) \to \infty$:
\begin{equation}
  \Phi_m(T, D) \simeq \varphi_0 \left(\frac{D}{\xi_0^+} \right)^{-\beta/\nu}
    \left[ P_2\mleft(\frac{1}{2}, 0\mright) + C \left(\frac{D}{\xi_b}\right)^{1/\nu} \right] \,,
\end{equation}
with some constant $C > 0$.
In particular, $\Phi_m(T,D)$, and thus the excess adsorption $\Gamma(T,D)$, do not diverge near the bulk critical point, which is a consequence of the OP conservation and at variance to the behaviour in the grand canonical ensemble \cite{dietrich1995} [\cref{critical-adsorption}].
Rather, for fixed $D$, the excess adsorption decreases linearly from its critical value at $T_c$ as the temperature is increased ($\xi_b(T) \gg D$):
\begin{equation}
  \Gamma(T \downarrow T_c, D) + \Gamma(T_c, D) \sim \xi_b^{-1/\nu} \sim T - T_c \,.
  \label{excess-scaling}
\end{equation}

\section{Simulation model and method}

For the MD simulations, we use a symmetric binary liquid with well-known bulk properties \cite{das2003, das2006jcp, roy2011epl, roy2016, pathania2021ats}.
Specifically, we consider an equimolar mixture of two components $A$ and $B$ which interact symmetrically with each other via the Lennard-Jones (LJ) pair potential,
\begin{equation}
  u_\text{LJ}(r; \varepsilon,\sigma)=4\varepsilon [(\sigma/r)^{12}-(\sigma/r)^6] \,,
\end{equation}
with parameters $\varepsilon$ and $\sigma$.
The potential is shifted and smoothly truncated at a cut-off distance $r_c=2.5\sigma$, which renders very good numerical stability with respect to energy conservation \cite{voigtmann2009, zausch2010, colberg2011}:
\begin{equation}
\label{particlepotential}
  U_{\mu\nu}(r)
  = \bigl[u_\text{LJ}(r; \varepsilon_{\mu\nu}, \sigma)-u_\text{LJ}(r_c; \varepsilon_{\mu\nu},\sigma)\bigr] \,
  {f((r-r_c)/h)}
\end{equation}
for $\mu, \nu \in \{A, B\}$, $h=0.005\sigma$, and $f(\zeta \leq 0)=\zeta^4/\bigl(1+\zeta^4\bigr)$ and $f(\zeta >0)=0$.
The interaction range $\sigma$ is the same for all pairs $\mu\nu$, whereas the interaction strengths are chosen as $\varepsilon_{AA}=\varepsilon_{BB}=2\varepsilon_{AB}=:\varepsilon$.
For the units of length and energy, we use $\sigma$ and $\varepsilon$, respectively, which implies the reduced temperature $T^*:=k_B T/\varepsilon$.
Of each species, we simulate $N_\mu$ particles such that species $\mu$ has concentration $x_\mu=N_\mu/N$ and such that the total number density is fixed at $\rho_b = N/V = 0.8\sigma^3$,
in terms of $N=N_A + N_B$ and the pore volume $V$.
We have previously determined the demixing phase diagram of this fluid
\footnote{We note that a force-shifted potential was employed in the studies in Refs.~\citenum{das2003, pathania2021ats}, which yields different values for non-universal properties, i.e., the critical temperature and the critical amplitudes.}
and the relevant critical behaviour in bulk \cite{roy2016}; in particular, the critical point was found to be
in the liquid phase at temperature $T_c^* = 1.629\pm0.001$ and concentrations $x_{A,c}=x_{B,c}=1/2$;
the non-universal amplitude of the correlation length $\xi_b(T)$ was obtained as $\xi_0^+ = (0.47 \pm 0.02)\sigma$,
and we have estimated for bulk OP amplitude $\varphi_0 = 0.77 \pm 0.02$.

The liquid is confined in between two planar walls separated by a distance $D$.
The area of each wall is $L\times L$ and is set by the lateral dimensions of the simulation box, with periodic boundary conditions applied along the two directions parallel to the walls;
the volume of the slit pore is thus $V=L^2 D$.
The walls are located at nominal positions $z=0$ and $z=D$, with the $z$-axis of the Cartesian coordinates chosen normal to the walls.
The surface--fluid interaction is the same for both walls and is given by the LJ wall potential (see, e.g., Refs.~\citenum{das2006sdsd,das2006pre,royepje2015}):
\begin{equation}\label{wallpotential}
  U_{\text{sf}}^{(\mu)}(\tilde z)=\frac{2\pi \varepsilon_w}{3} \left[
    \frac{2}{15}\left(\frac{\sigma}{\tilde z}\right)^9
    - w_\mu \left(\frac{\sigma}{\tilde z}\right)^3 \right],
\end{equation}
in terms of the distances $\tilde z=z+\sigma/2$ and $\tilde z=D+\sigma/2 - z$ to each of the walls.
The offset $\sigma/2$ is introduced in the surface potential to better match the volume that is effectively available to the fluid particles with the nominal pore volume $L^2 D$.
The form of the potential results from an integral over the LJ pair potential, modelling the wall as an infinite half-space which is filled homogeneously with LJ particles.
The surface potential $U_{\text{sf}}^{(\mu)}(\tilde z)$ is further shifted and smoothly truncated at the cutoff distance $r_{c,\text{sf}}=2.5\sigma$ as described above for the pair interaction.
The energy $\varepsilon_w$ and the dimensionless parameter $w_\mu$ control the
overall attraction strength and the relative surface preference for one of the two fluid components, respectively;
$U_{\text{sf}}^{(\mu)}(\tilde z)$ has its minimum at $\tilde z_\text{min} = (2/5w_\mu)^{1/6}\sigma$, for $w_\mu > 0$,
with depth $(2\pi \varepsilon_w/9)\sqrt{10w_\mu^3}$.
We use the same value $\varepsilon_w$ for both walls of the pore, and we put $w_A=1$ and $w_B=0$, which amounts to symmetric boundary conditions with a preference for $A$ particles.

All results presented here are obtained at or above the bulk critical temperature of the binary liquid, $T \geq T_c$ (after a quench from the mixed phase at $T > T_c$), and at the critical concentrations $x_A=x_B=1/2$,
i.e., for zero total OP, $\phi_\text{tot} = x_A-x_B=0$.
Extensive MD simulations were carried out in the canonical ensemble with the Nosé--Hoover thermostat
(NHT) chain \cite{martyna1992}, using an integration timestep of $\delta t=0.001t_0$;
the unit of time is $t_0=\sqrt{m\sigma^2/\varepsilon}$ in terms of the particle mass $m$, which is the same for $A$ and $B$ particles.
To mitigate the computational issues associated with the simulation of critical fluids, such as critical slowing down \cite{folk2006} and finite-size corrections \cite{binder1981, roy2013, roy2014, pathania2021ats}, we have used large system sizes for up to 
\num{300000} particles over a long time span of \SI{e4}{t_0};
in addition, ensemble averages were taken over up to 20 independent initial configurations for each parameter set.
Specifically, we have used large simulation boxes of lateral length up to $L=100\sigma$ and we have varied the pore width from $D=10\sigma$ to $30\sigma$.
The simulations were performed with the software \emph{HAL's MD package} \cite{colberg2011,HALMD}, which exploits the high degree of parallelism of recent accelerator hardware \cite{skoblin2023}, e.g., graphics processors (GPUs), and which is an efficient and precise tool for the study of inhomogeneous fluids \cite{hoefling2015, hoefling2024jcp, chaudhuri2016prb, ebrahimi-viand2020jcp}.
For example, a simulation run for $D=20\sigma$ over a period of $10^4 t_0$  (\num{e7} steps),
takes about \SI{1}{h} on a single GPU of type Nvidia A40 
and about \SI{5.8}{h} on the much older hardware generation, Nvidia Tesla K20Xm. 
%
%

The simulation data were stored in the structured, compressed, and portable H5MD file format \cite{debuyl2014}.
Laterally averaged number density profiles $\rho_\mu(z) = N_\mu(z) / (L^2 \Delta z)$ were calculated in a post-processing step by counting the number $N_\mu(z)$ of particles of each species $\mu$ that are located in the interval $[z - \Delta z/2, z + \Delta z/2)$ along the $z$-axis for a bin width of $\Delta z = 0.5\sigma$.
For the total density profile $\rho(z)$, shown in the inset of \cref{fig:op1}(b), we used a different technique to be described elsewhere, which is based on Fourier transforms, avoids the spatial binning and reduces the amount of data to be stored. The method includes a slight smoothing of the obtained profile $\rho(z)$, here with a Gaussian filter of width $\sigma/8$.

\section{Results and Discussion}

\begin{figure}
\centering
\includegraphics{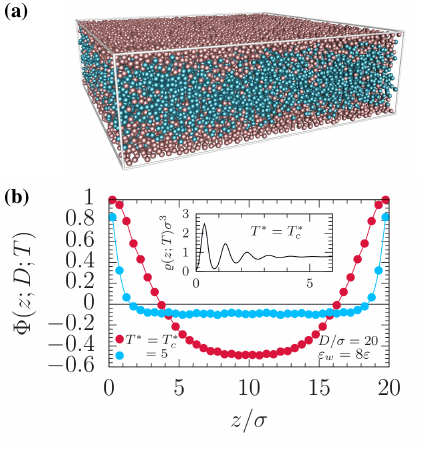}
\caption{(a)~Exemplary configuration of an equimolar, critical binary liquid confined to a slit pore. Brownish and turquoise beads show the particle positions of the fluid components A and B, respectively, in a snapshot of the MD simulations for lateral edge length $L=50\sigma$ and pore width $D=15\sigma$.
Both wall surfaces (at the top and bottom of the wire frame, not shown) have the same strong preference for component~A.
The binary mixture is at its consolute point, and the mass of each component is locally conserved within the simulations, i.e., the order parameter (OP) for demixing is a conserved field.
~(b)~Excess adsorption of the laterally averaged, local OP $\Phi(z)$ for a high temperature in the mixed phase ($T^*=5$, light blue symbols) and at the bulk consolute point of the mixture, $T^*=T_c^*\approx 1.63$ (red symbols).
The wall surfaces are located at effective positions $z=0$ and $z=D = 20\sigma$.
Filled symbols are simulation data and solid lines smoothly connect the data. 
The inset shows the corresponding number density profile $\rho(z)$ of the critical fluid;
the mean density is $\rho_b = 0.8\sigma^{-3}$.
}
\label{fig:op1}
\end{figure}

\begin{figure}
\centering
\includegraphics[width=\figwidth]{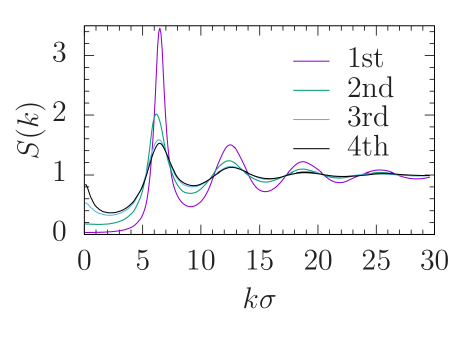}
\caption{Local, quasi-twodimensional density--density structure factor $S_{\rho\rho}(|\vec k|)$ calculated separately for the first four layers of particles closest to the pore surface (see $\rho(z)$ in \cref{fig:op1}); the wave vector $\vec k$ points parallel to the walls \cite{hoefling2020jcp}.}
\label{fig:ssf}
\end{figure}

Within the MD simulations, we have obtained the surface adsorption profiles $\Phi(z)$ of binary liquids
confined to a slit pore which preferentially attracts component $A$ at both of its walls with equal strength [\cref{fig:op1}(a)];
unless stated differently, we consider a strong surface attraction, $\varepsilon_w=8\varepsilon$ [see \cref{wallpotential}].
The present study is restricted further to binary liquids at critical composition ($x_A = x_B = 1/2$) and in the mixed phase ($T \geq T_c$), so that the temperature $T$ controls the distance to the bulk critical point of the demixing transition.
$\Phi(z)$ serves also as the corresponding local OP, it is defined such that positive values indicate an excess of $A$ particles.
Since the set-up exhibits a mirror symmetry, $\Phi(z; T, D) = \Phi(D-z; T, D)$, with respect to the central plane ($z=D/2$), we will  often discuss merely the behaviour near one of the surfaces.

Two exemplary OP profiles $\Phi(z)$ for two temperatures, in the mixed phase and at criticality, are shown in \cref{fig:op1}(a).
At the higher temperature, $T^*=5 \approx 3 T_c^*$, the profile is comparably flat in a large part of the pore but increases sharply upon approaching one of the surfaces (e.g., $z \to 0$), the latter reflecting the expected surface enrichment of A particles.
Upon decreasing the temperature to $T_c$, the contact value $\Phi(z \to 0)$ increases to its largest possible value, $\Phi(z \to 0) = 1$, which we attribute to the choice of a strong surface interaction ($\varepsilon_w=8\varepsilon$);
this value of $\Phi(z)$ corresponds to a pure layer of A particles next to the surfaces.
Concomitantly, the OP profile broadens as $T \downarrow T_c$, reflecting the emergence of critical fluctutations.
However, the width of the surface enrichment layer remains finite (of the order of $3{-}4\sigma$) despite the bulk correlation length $\xi_b(T)$ diverging at $T_c$, which is a consequence of the confinement to a slit and of the aforementioned symmetry of $\Phi(z)$ at $z=D/2 = 10\sigma$.

\begin{figure}
\centering
\includegraphics[width=\figwidth]{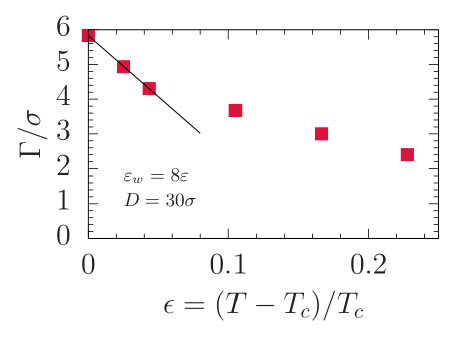}
\caption{Total excess adsorption $\Gamma$ as function of the reduced temperature $\epsilon = (T - T_c)/T_c$, defined relative to the critical value $T_c$.
The excess adsorption is proportional to the OP value in the middle of the slit pore, $\Gamma=-\Phi_m D/2$ with $\Phi_m = \Phi(z=D/2)$, due to the use of the canonical ensemble with conserved OP [\cref{phi-m}].
The binary liquid is at critical composition such that the critical point is approached from the mixed phase upon $\epsilon \downarrow 0$.
For this case, the solid line tests an asymptotically linear dependence of $\Gamma$ on $\epsilon$ [\cref{excess-scaling}].
The simulation data (symbols) were obtained for fixed pore geometry ($D=30\sigma$, $L = 100\sigma$) and fixed surface interaction strength $\epsilon_w = 8 \epsilon$; the statistical uncertainties are smaller than the size of the symbols.
}\label{fig:op-middle}
\end{figure}

Furthermore, the conservation of the OP implies that the integrated OP $\phi_\text{tot}$ does not change upon varying the temperatures; here, we use $\phi_\text{tot}=0$.
It follows from \cref{op-tot} that the OP value in the middle of the slit pore must be negative, $\Phi_m = \Phi(z=D/2) < 0$, to compensate the surface enrichment.
This is a specific feature of the canonical ensemble and unlike in the grand canonical ensemble, in which the local OP remains positive, $\Phi(z) \geq 0$, for symmetric surfaces both preferentially attracting component~A.
For conserved OP, the value $\Phi_m$ thus serves as a measure of the total excess adsorption $\Gamma(T,D) = -\Phi_m(T,D) \, D /2$ [\cref{phi-m}].
The latter is predicted to vary linearly with $T$ near the critical temperature [\cref{excess-scaling}],
which is corroborated by our simulation data (\cref{fig:op-middle}).

In contrast to the monotone decay of $\Phi(z)$ for $0 \leq z \leq D/2$ (\cref{fig:op1}), the number density of the critical fluid ($T=T_c$) exhibits a pronounced layering near the surfaces (inset of \cref{fig:op1}):
the density profile $\rho(z)$ shows rapidly decaying oscillations around the mean density $\rho_b=0.8\sigma^{-3}$, which have a large amplitude next to the surface at $z=0$ and which are barely visible in the middle of the pore, for $z \gtrsim 4\sigma$; the period of the oscillations is of the order of the molecular length $\sigma$.
Such oscillations hinder the scaling analysis of critical surface profiles, and their absence in the present OP profiles (i.e., in the concentration and thus the excess adsorption) underscores that, in this respect, binary liquids are favourable systems for the study of critical behaviour.
We note that the critical behaviour of the structural and dynamic properties of binary liquids is governed by two fluctuating fields, viz., the total number density and the composition.
The investigated binary liquid (with $\rho_b=0.8\sigma^3$) was demonstrated \cite{roy2016} to be
sufficiently far away from its liquid--vapor critical point and thus the density field
$\rho(z)$ acts as a secondary field \cite{folk2006}, which does not exhibit critical enhancement
(see, e.g., the bulk structure factors in Figs. 4 and 5 of Ref.~\citenum{roy2016}).

In order to further elucidate the lateral structure of the layers near the surface, we have calculated the lateral structure factor $S_{\rho\rho}(|\vec k|)$ of the density for each layer separately, by restricting the calculation to quasi-twodimensional slabs encompassing one layer each and such that the wave numbers point parallel to the wall surfaces, for details see Ref.~\citenum{hoefling2020jcp}.
Upon approaching the surface, layer by layer, the apparent compressibility $\propto S_{\rho\rho}(0)$ is reduced and the degree of local ordering increases, as follows from the growth and the sharpening of the first peak of $S_{\rho\rho}(|\vec k|)$ (\cref{fig:ssf}).
However, also the first fluid layer, which is in direct contact with the wall, does not exhibit crystallisation but remains in a fluid state.

\begin{figure}
\centering
\includegraphics[scale=1.2]{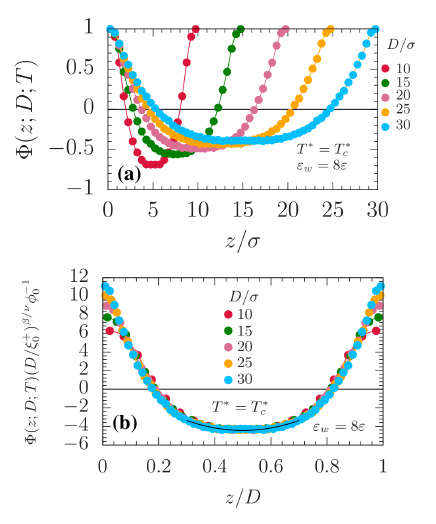}
\caption{(a) Surface adsorption profiles of the critical mixture confined to slit pores of five different widths $D$ and
constant aspect ratio $L/D=3.3$.
Symbols show simulation data for the laterally averaged OP profiles $\Phi(z)$, and all results correspond to $T^*=T_c^*$ and $\varepsilon_w=8\varepsilon$.
(b) Data collapse of the rescaled OP profiles of panel~(a) onto the scaling function
$P_2(\bar z, 0)=\varphi_0^{-1} (D/\xi_0^+)^{\beta/\nu} \Phi(z;T,D)$
in terms of the the rescaled distance $\bar z = z/D$ [\cref{op1}].
The black solid line indicates a fit to a parabolic profile in the central part of the pore [cf. \cref{op-cosh-profile}].
}
\label{op_scaling_Tc}
\end{figure}

Next, we investigate the dependence of the OP profile $\Phi(z)$ of the critical mixture on the pore width $D$ for fixed temperature $T=T_c$.
For strong surface attraction ($\varepsilon_w=8\varepsilon$), the contact value $\Phi(z \to 0)$ at the surface does not
exhibit any prominent $D$ dependence [\cref{op_scaling_Tc}(a)]. 
As $D$ is increased, the profile becomes deeper ($|\Phi_m|$ increases) and broadens: the thickness of the
surface adsorption layer (e.g., the distance to the surface at which $\Phi(z)$ changes its sign) increases.

In order to test the scaling property of these critical OP profiles, we have plotted the combination
$\varphi_0^{-1} (D/\xi_0^+)^{\beta/\nu} \Phi(z,T,D)$ as a function of the scaling variable $\bar z = z/D$
for a range of pore widths $D$ that is spread out by a factor of 3. 
\Cref{op_scaling_Tc}(b) demonstrates convincingly that this rescaling of the OP profiles of \cref{op_scaling_Tc}(a)  yields data collapse onto the scaling function
$P_2(\hat z, 0)$; due to $T=T_c$, we have $\hat D = D/\xi_b(T_c) = 0$.
Deviations from this universal master curve occur for distances $z \lesssim 0.8\sigma$, which corresponds to the non-universal microscopic regime.
We note that, in order to avoid spurious finite-size corrections to the scaling, we found it necessary to also fix the aspect ratio $L/D=3.3$ of the pore geometry; the ratio $L/D$ enters the scaling form of the OP profile [\cref{op1}] as an additional finite-size scaling variable (see, e.g., Refs.~\citenum{binder1981} and \citenum{pathania2021ats}).

\begin{figure}
\centering
\includegraphics[scale=1.2]{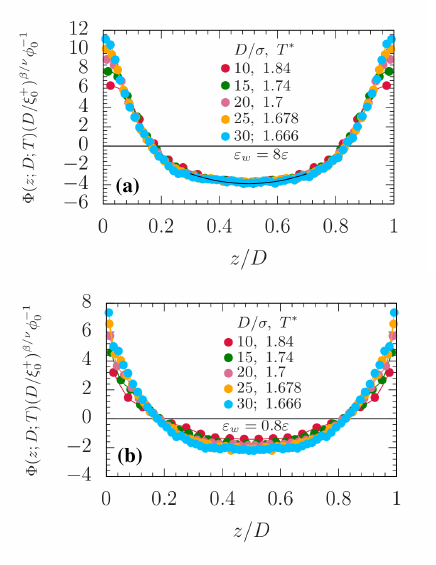}
\caption{%
Scaling of the surface adsorption profiles of near-critical binary liquids confined to five different slit pores for (a)~strong ($\varepsilon_w=8\varepsilon$) and (b)~weak ($\varepsilon_w=0.8\varepsilon$) surface interactions.
The plots test the data collapse onto the scaling function
$P_2(\bar z, \hat D) = \varphi_0^{-1} (D/\xi_0^+)^{\beta/\nu} \Phi(z;T,D)$
as function of $\bar z = z/D$ [\cref{op1}].
The data (symbols) correspond to different combinations of pore width $D$ and temperature $T$, the latter entailing the bulk correlation length $\xi_b(T)$, such that $\hat D = D/\xi_b(T) \simeq 5.9$ is constant;
the aspect ratio of the pore is kept fixed at $L/D = 3.3$.
}
\label{fig:op_scaling_fixed_D_xi}
\end{figure}

The scaling function $P_2$ of the OP profile $\Phi(z;T,D)$ depends on two scaling variables, $\bar z = z/D$ and $\hat D = D/\xi_b(T)$ [\cref{op1}].
For elevated temperatures, $T > T_c$, a similar data collapse as for the critical mixture can be obtained by considering a range of combinations of pore width $D$
and temperature $T$ such that $\hat D = D/\xi_b(T) = (D/\xi_0^+)\epsilon^{\nu}$ is kept constant.
Indeed, the rescaled data for the OP profiles with $\hat D = 5.9$ and for strong surface attraction ($\varepsilon_w=8\varepsilon$) collapse onto the master curve $P_2(\bar z, \hat D = 5.9)$ similarly well as in \cref{op_scaling_Tc}(b) for $T=T_c$.
However, repeating this analysis for weak surface attraction ($\varepsilon_w=0.8\varepsilon$), the quality of the data collapse is considerably worse [\cref{fig:op_scaling_fixed_D_xi}(b)]. In this situation, the OP profiles depend also on the strength $\varepsilon_w$ of the surface field, which enters the scaling function in \cref{op1} as another scaling variable, in addition to $z/D$, $D/\xi_b$, and $L/D$ (see, e.g., Ref.~\citenum{gross2016} for details).

\begin{figure}
\centering
\includegraphics{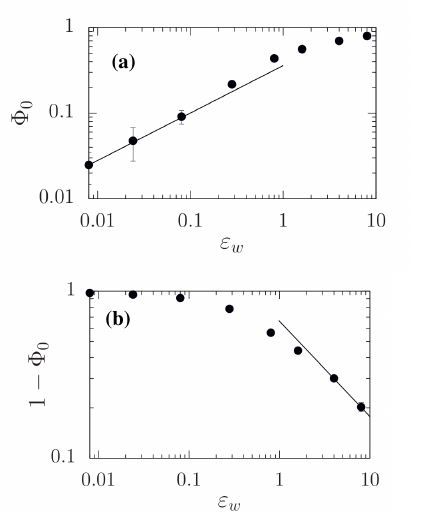}
\caption{Contact value $\Phi_0$ of the surface adsorption of the critical mixture ($T= T_c$) as function of the
surface interaction strength $\varepsilon_w$ on double-logarithmic scales.
Both panels show the same simulation data (symbols), corresponding to $D=20\sigma$ and $L=100\sigma$,
with panel~(b) testing the approach to the maximum OP value, $\Phi_0 \leq 1$.
Asymptotic power laws (solid lines) are fits to the data for the regimes of (a) weak surface interaction,
$\Phi_0(\epsilon_w \to 0) \sim \epsilon_w^{0.55}$,
and (b) strong surface interaction, $1-\Phi_0(\epsilon_w \to \infty) \sim \epsilon_w^{-0.57}$.
}
\label{fig:op-ew}
\end{figure}

For a more systematic investigation of the dependence of the OP profile on the strength $\varepsilon_w$ of the surface attraction, we consider the OP value close to the surface.
Within the mesoscopic GL theory employing a surface field $h_1$ \cite{gross2016}, it was predicted that $\Phi(z \to 0)=\Phi_0$ displays a crossover from the scaling $\Phi_0 \sim h_1$ for weak fields, $h_1 \to 0$,
to $\Phi_0 \sim h_1^{1/2}$ for strong fields, $h_1 \to \infty$.
Since there is no direct mapping between the surface field $h_1$ of GL theory and the surface interaction parameters ($\varepsilon_w, w_A$) used in the MD simulations, we check if the scaling of $\Phi_0$ with $\varepsilon_w$ is analogous to what has been predicted within GL theory for the dependence on $h_1$;
to this end, the mean-field exponents should be replaced by their counterparts in the 3D Ising universality class.
However, the contact value $\Phi(z\to 0)$ appears to be not suited for this task since universality of the surface adsorption profiles is expected to hold only for $z \gg \sigma$.
Hence, we define the surface value $\Phi_0$ as an integral over the particle layer second-closest to the surface:
\begin{equation}\label{op0}
\Phi_0 = \frac{1}{0.5\sigma} \int_{\sigma}^{1.5\sigma} \Phi(z) \,\diff z \,.
\end{equation}
Using this definition of $\Phi_0$, the data from MD simulations of the critical mixture ($T=T_c$, $D=20\sigma$, and $L=100\sigma$)
are described the behaviour $\Phi_0(\varepsilon_w \to 0) \sim \varepsilon_w^{0.55}$ for weak surface interaction strengths, varying from $\varepsilon_w=\num{e-2} \varepsilon$ to $\approx 0.2 \varepsilon$ [\cref{fig:op-ew}(a)].
Despite these low values of $\varepsilon_w$, the observed scaling is at variance with the expectation from perturbation theory for Ising and GL models that $\Phi_0$ scales linearly with the surface field \cite{binder2003jsp,gross2016}.
For the the strong-surface field behaviour, one needs to account for the bound $|\Phi_0| \leq 1$ due to the definition of $\Phi(z)$, similarly as above for the excess adsorption $\Gamma$. Thus, we consider $1-\Phi_0$ in order to test the approach of $\Phi_0$ to its saturation value.
Fitting a power law to the data points for the two largest surface interactions simulated ($\varepsilon_w = 4\varepsilon$ and $8\varepsilon$) suggests that
$1-\Phi_0(\varepsilon_w \to \infty) \sim \varepsilon_w^{-0.57}$ [\cref{fig:op-ew}(b)].
We note that this value of the exponent is larger than the standard surface exponent \cite{binder2003jsp,gross2016} $\Delta_1 \approx 0.46$ and that, to the best of our knowledge, a theoretical prediction for the strong-surface field behaviour of $\Phi_0$ under the simulated conditions is not available.

\section{Summary and conclusions}

Understanding the surface critical behaviour of a fluid confined to a slit pore is demanding due to the presence of multiple length scales: there are the molecular size $\sigma$ and the pore width $D$, the temperature dependence implies the bulk correlation length $\xi_b(T)$,
the strength of the surface interaction $\varepsilon_w$ adds another implicit length, and, in simulations, the finite lateral extent $L$ of the slit pore cannot be ignored.
Here, we have used large-scale, off-lattice MD simulations to explore the scaling of the excess adsorption profiles $\Phi(z)$ of a binary liquid with locally conserved mass and concentration;
the liquid is kept confined to a slit whose walls have selective surface adsorption preferences.
For the demixing transition, the excess adsorption plays the role of the OP field, which develops critical fluctuations near the liquid--liquid critical point.
Relying on the corresponding scaling form [\cref{op1}],
we have shown that the data for laterally averaged OP profile $\Phi(z;T, D)$ for a range of temperatures $T$ and pore widths $D$ collapse onto a master curve upon suitable rescaling (\cref{op_scaling_Tc}).
This master curve is given by the scaling function $P_2$ in the limit of strong surface adsorption and for fixed ratios $D/\xi_b$ and $L/D$.
The obtained OP profiles behave monotonically near the surfaces, also in the presence of a pronounced layering in the number density profiles [\cref{fig:op1}(b)].
Furthermore, our data suggest that OP profiles obtained within the MD simulations are better described by the regime of strong surface attraction, even for low values of $\varepsilon_w$, which raises a question on the precise mapping between the particle model used in the MD simulations and GL theory (\cref{fig:op-ew}).
Overall, our results are in qualitative agreement with mean-field predictions from field-theoretical calculations \cite{gross2016,gross2017,rohwer2019pre}.

Whereas our study is based on the generic and somewhat artificial Lennard-Jones pair interaction,
we anticipate that our findings for critical adsorption are universal and can be transferred \emph{mutatis mutandis} to real-world binary liquids, including water--lutidine mixtures.
In particular,
understanding the influence of surfaces on a confined fluid is important for the study of critical Casimir forces \cite{fukuto2005prl, rafai2007physa, hertlein2008, dantchev2023pr, Gambassi:SM2024, Wang:NC2024}. We hope this work will motivate future work on the non-equilibrium dynamics of near-critical confined fluids and on the dynamic response to a temperature quench \cite{sengers2013casimir, furukawa2013, rohwer2019njp}.

\begin{acknowledgments}
 We thank Siegfried Dietrich and Markus Gross for helpful discussions.
 Financial support by Freie Universität Berlin via a research stipend under the Future Development Concept: ``International Network University'' (S.R.)
 and by Deutsche Forschungsgemeinschaft (DFG, German Research Foundation) under Project No.\ 523950429 (F.H.) is gratefully acknowledged.
\end{acknowledgments}

\bibliography{critical_fluids}

\end{document}